\documentclass[preprint,12pt,authoryear]{elsarticle}

\usepackage{amssymb}
\usepackage{amsmath}

\graphicspath{{pix/}}
\usepackage{hyperref}

\journal{The Journal of Academic Librarianship}

\begin{document}

\begin{frontmatter}

\title{Cloud-based digitization workflow with rich metadata acquisition for cultural heritage objects} %

\author[1]{Krzysztof Kutt\corref{cor1}}
\ead{krzysztof.kutt@uj.edu.pl}
\ead[url]{https://krzysztof.kutt.pl/}

\author[1]{Luiz do Valle Miranda}
\ead{luiz.miranda@uj.edu.pl}

\author[2]{Jakub Gomu\l{}ka}
\ead{jgomulka@agh.edu.pl}

\author[1]{Grzegorz J. Nalepa}
\ead{grzegorz.j.nalepa@uj.edu.pl}
\ead[url]{https://gjn.re/}

\cortext[cor1]{Corresponding Author}

\affiliation[1]{organization={Department of Human-Centered Artificial Intelligence, Institute of Applied Computer Science, Faculty of Physics, Astronomy and Applied Computer Science, Jagiellonian University},
            addressline={S. \L{}ojasiewicza 11}, 
            city={Krak\'ow},
            postcode={30-348}, 
            country={Poland}}

\affiliation[2]{organization={Faculty of Humanities, AGH University of Krakow},
            addressline={Czarnowiejska 36}, 
            city={Krak\'ow},
            postcode={30-054}, 
            country={Poland}}

\begin{abstract}
In response to several cultural heritage initiatives at the Jagiellonian University, we developed a new digitization workflow in collaboration with the Jagiellonian Library (JL).
The solution is based on easy-to-access technological solutions---Microsoft 365 cloud with MS Excel files as metadata acquisition interfaces, Office Script for validation, and MS Sharepoint for storage---that allows metadata acquisition by domain experts %
regardless of their experience with information systems.
The ultimate goal is to create a knowledge graph that describes the analyzed collections, linked to general knowledge bases, as well as to other cultural heritage collections, so careful attention is paid to the high accuracy of metadata and proper links to external sources.
The workflow was evaluated in two pilot studies and in two workshops, %
which allowed for its refinement and confirmation of its correctness and usability for JL.
The knowledge graph created as a result of these pilot studies was made available in a public git repository.
As the proposed workflow does not interfere with existing systems or domain guidelines regarding digitization and basic metadata collection in a given institution, %
but extends them in order to enable rich metadata collection, not previously possible, we believe that it could be of interest to all GLAMs. %
\end{abstract}

\begin{keyword}
Cultural heritage \sep Metadata \sep Knowledge graphs \sep GLAM \sep Digitization \sep Spreadsheets %

\end{keyword}

\end{frontmatter}

\section{Introduction and motivation}
\label{sec:intro}

Jagiellonian University (JU), the oldest Polish university and one of the oldest universities in continuous operation in the world, holds in its collections a number of cultural heritage objects, including medieval manuscripts, social life documents representing a wide range of historical periods, memorabilia of former employees and students of the university, and a collection of authentic relics from the circle of ancient Mediterranean cultures.
These collections have attracted greater research interest recently due to large cultural heritage initiatives at the JU, particularly the so-called flagship projects (FPs)\footnote{See \url{https://id.uj.edu.pl/en_GB/projekty-flagowe} for more details.} aimed at interdisciplinary and international cooperation to enrich the JU's research ecosystem. Five of the 21 FPs are focused on cultural heritage research, including digitization, standardization of descriptions, integration of collections, and more detailed studies of selected items (cf. Sect.~\ref{sec:ju}).

As a team of computer scientists and knowledge engineers, we are actively collaborating with these cultural heritage initiatives to assist researchers---philologists, historians, philosophers, librarians, archivists, curators, etc.---by developing proper tools for metadata acquisition during the digitization process, which will ultimately lead to the creation of knowledge base(s) that describe the collections under analysis.

With this in mind, we decided to take advantage of knowledge graphs~\citep{hogan2021graphs} and ground the entire solution in the Semantic Web technologies~\citep{hitzler2021review} and the linked data approach~\citep{heath2011linked,hyvonen2019knowledge}.
Adopting this approach will allow for the explicit definition of the meaning of individual metadata, the preservation of a flexible data model that can be easily adapted to the needs of individual collections or research objectives, the efficient integration of data across disciplines and organizations, and will facilitate future data processing in automated systems~\citep{janowicz2015train}.

During the development of metadata acquisition tools, a number of challenges and requirements had to be addressed, among which the most important were
the tight time frame for tool development,
the lack of resources to develop dedicated software,
the need to use tools that are as familiar and accessible as possible to domain experts,
and the scalability of the solution to collections with hundreds of thousands of objects.
Taking into account all of these, we decided to base the solution on Microsoft 365 cloud and use
Microsoft Excel spreadsheets as an interface for metadata collection,
Office Script to facilitate the use of the spreadsheets,
Python scripts for initial metadata validation,
dedicated space in Microsoft Sharepoint for data storage,
and Python scripts for ontology population from gathered data.
The whole solution is complemented by a dedicated workflow based on the developed tools.

The original contribution of the paper is the prototype of a digitization workflow, based on easy-to-access technological solutions (Microsoft 365 cloud), that allows rich metadata acquisition by domain experts regardless of their experience with information systems, resulting in the creation of knowledge base filled with gathered metadata.
The proposed solution does not interfere with existing systems or domain guidelines regarding digitization and basic metadata collection in a given institution (e.g., file type, image quality, use of Dublin Core/MARC 21), but extends them in order to enable rich metadata collection, not previously possible.
Therefore, we believe that it could be of interest to all GLAMs (galleries, libraries, archives, and museums).
While the use of Excel for metadata collection is quite common, basing the entire digitization workflow on Microsoft 365 cloud to the best of our knowledge has not been described before.

The remainder of the paper is structured as follows.
Sect.~\ref{sec:ju} introduces selected cultural heritage collections and digitization projects at Jagiellonian University.
The related work is summarized in Sect.~\ref{sec:related}.
The requirements for the entire workflow are listed in Sect.~\ref{sec:goal}.
The tools used in the proposed solution are outlined in Sect.~\ref{sec:methods},
while their evaluation during actual digitization workflow resulting in the first version of knowledge base is described in Sect.~\ref{sec:workflow}.
Finally, Sect.~\ref{sec:conclusions} concludes the paper.

\section{Cultural heritage collections and digitization landscape at Jagiellonian University}
\label{sec:ju}

From the point of view of cultural heritage collections, the Jagiellonian Library (JL) should be considered a central part of the Jagiellonian University.
Although it was not founded together with the University in 1364, the Library is a successor to various university libraries and collections that have existed since the 15th century.
For many years, it also played the role of a national library, the so-called \emph{bibliotheca patria}~\citep{bakowska2005jl}.
Now, JL holds 6,798,156 items\footnote{For official numbers, see: \url{https://bj.uj.edu.pl/en_GB/about-the-library/mission-history-and-collections/collections/jagiellonian-university-library-system-holdings-in-numbers}.} in its collections (not including items stored in faculty libraries), with many cultural heritage treasures including 11th century documents, collection of incunabula or Dutch, Italian and French engravings~\citep{bakowska2005jl}.

Jagiellonian University's heritage collections are also spread in many places outside the Jagiellonian Library.
For example, documents and photos of university staff and students are kept in the JU Archive,
the JU Museum keeps a collection of memorabilia,
the mediterranean and prehistoric archaeological collections are kept at the JU Institute of Archaeology~\citep{wozny2018archeological}.
There are also the photo collection of the Institute of Art History,
the Ignacy Jan Paderewski Center for the Documentation of 19th- and 20th-century Polish Music,
and many other collections~\citep{zieba2020museum}.

Most of the aforementioned cultural heritage objects have not yet been digitized.
To illustrate, JL has digitized and made available to the public 904,624 objects.
Considering the special collections alone---that is, excluding collections of books and periodicals, many of which cannot be digitized due to legal restrictions---comprising 2,390,834 objects, this means that more than 60 percent of the collections are still not digitized.

These collections have attracted greater research interest recently due to large cultural heritage initiatives at the JU, particularly the so-called flagship projects (FPs)\footnote{See \url{https://id.uj.edu.pl/en_GB/projekty-flagowe} for more details.} aimed at interdisciplinary and international cooperation to enrich the JU's research ecosystem. Five of the 21 FPs are focused on cultural heritage research, including digitization, standardization of descriptions, integration of collections, and more detailed studies of selected items. %
This
cultural heritage landscape is accompanied by various systems used to store metadata about objects:
\begin{enumerate}
    \item JL uses the Alma interlibrary catalog (\url{https://katalogi.uj.edu.pl/}), which stores data in MARC 21 format, and the Jagiellonian Digital Library (\url{https://jbc.bj.uj.edu.pl/}) based on the dLibra system, which stores scanned texts in PDF format along with their metadata in the Dublin Core standard.  
    \item The JU Museum stores descriptions of many artifacts collected by various university units in the MuzUJ system (\url{https://muzuj.uj.edu.pl/}). This metadata is richer than the basic MARC 21/Dublin Core description, but is not based on shared vocabularies or ontologies. Currently, the MuzUJ system does not have a public interface -- interested persons send a request to Museum staff, who perform searches in the system.
    \item There are also many other small systems and databases that store information about individual collections, for example, the photo collection of the Institute of Art History is presented in \url{http://www.fototeka.ihs.uj.edu.pl/}.
\end{enumerate}

\section{Related work}
\label{sec:related}

A common starting point for the decision of what software to include in the digitization workflow for metadata acquisition is evaluating existing digital library softwares (DLS). This evaluation consists of a mapping of institutional requirements to features offered by different tools. Among institutional requirements are the target format for metadata, the extent to which an institution is open for changes in their existing procedures and the training capabilities for introducing new interfaces for those performing the metadata entry operations. Among features of DLS are the supported metadata standards, licensing model, user interface design for metadata entry,  controlled vocabularies, learning curve, and their capacity for future customization \citep{Mukherjee02072020}. 

As discussed above, the limitations of the current digitization workflow of the Jagiellonian Library restricts the manual metadata entry process and storage to the use of tools present in the Microsoft 365 cloud package. Despite this restriction, an evaluation of existing DLS and a presentation of its current uses can still be useful for an outline of the target functionalities of the Microsoft cloud-based metadata acquisition workflow.

\citet{verma2018comparative} highlight three widespread DLS: DSpace\footnote{\url{https://www.dspace.org/}}, Greenstone\footnote{\url{http://www.greenstone.org/}}, and ePrints\footnote{\url{https://www.eprints.org/}}. DSpace is an open-source digital repository platform designed to capture, store, manage, and share digital content, widely used by academic institutions, libraries, and organizations for preserving scholarly outputs and institutional assets. It supports customizable metadata standards, interoperability protocols like OAI-PMH, and offers a user-friendly interface for managing collections. \citet{verma2018comparative} affirm that there are more than 1000 repositories which are using DSpace in at least one part of the development process. \citet{Chen2012} have reported on the process of using DSpace in the implementation of a Taiwan Digital Library History Library as both a metadata repository and a user-interface. \citet{Kurtz_2010} examines the influence of DSpace in The University of New Mexico, The University of Washington and Ohio State University as a metadata acquisition software and repository in the metadata quality of created records. From this brief investigation, DSpace appears to be a frequently chosen option due to its flexibility, scalability, and strong support for metadata management, including the use of linked data service and compatibility with discovery systems.

Greenstone is another open-source DLS that enables the creation, management, and dissemination of digital collections, and supports metadata formats including Dublin Core, as well as the OAI-PMH interoperability protocol. Greenstone provides a user interface for cataloging (i.e., inserting and editing records) and browsing digital collections, however it does not natively support linked data. A comprehensive overview of Greenstone current advancements can be found in \citep{bainbridge2020renewed}. A list of examples of libraries' usage of Greenstone can be found in \url{https://www.greenstone.org/examples}. Finally, ePrints is also a open-source DLS for creating and managing digital repositories, and includes customizable metadata formats and interoperability, exposition of its metadata in RDF, integration with third-party services such as ORCID and a user-friendly interface and strong search capabilities for both end users and administrators. \citet{Pramudyo2020eprints} present the use of ePrints in the Universitas Brawijaya Library as part of a process of metadata interoperability between several libraries using different digital library software.

An alternative software with a proprietary licensing model is CONTENTdm\footnote{\url{https://www.oclc.org/en/contentdm.html}}, a digital asset management software used by libraries to store, manage, and provide access to digitized collections that offers extensive support and hosting services for its users. \citet{payant2020contentdm} consists of a report that portrays the metadata fields, mappings, and resources used for creating and managing metadata in Utah State University's CONTENTdm digital collections. \citet{bahnemann2021transforming} share the findings of a pilot project investigating conversion methodologies for metadata originating from CONTENTdm into linked data.

The four aforementioned widespread DLS were developed and released in the late 1990s and early 2000s, that is, slightly before the publishing RDF 1.0 and OWL 1.0 specification. In the coming years, linked data became an important part of libraries' endeavor for making their records more widely available, findable and interoperable \citep{hallo2016digitallibrary}. While DSpace, ePrints and, to a lesser extent, CONTENTdm have provided updates for these tools supporting linked data solutions, other DLS have emerged with their functionalities based on the principles of linked data.

Omeka-S\footnote{\url{https://omeka.org/s/}} is an open-source platform for managing and sharing digital collections, including their metadata. It supports Linked Open Data with features like customizable metadata schemas, RDF mapping, and external LOD system interoperability. It possesses a user interface that allows users to manage, create, and display exhibits. In \citet{popovic2020omeka} one can read about the experience of implementing Omeka-S as data insertion, storage and interface at the Faculty of Mining and Geology Digital Repository at the University of Belgrade. Another example of usage of Omeka-S for semantic annotation, data storage and data browsing and querying can be found in \citet{bikakis2021omeka} for the Archives Henri Poincaré. Another LOD-based DLS is Arches-HIP\footnote{\url{https://arches-hip.readthedocs.io/en/latest/}}, an open-source platform for managing heritage data, emphasizing LOD principles for interoperability. It stands out due to its ability to integrate spatial analysis and visualization tools. In \citet{Yang2018Arches} Arches-HIP is used for the registration and integration  of Taiwan's cultural heritage metadata into CIDOC CRM-based ontologies and presentation  through a web platform in spatially meaningful representations. Finally, other tools used in in GLAM-related metadata tasks are, among others, CollectionSpace\footnote{\url{https://collectionspace.org/}}, Archivematica\footnote{\url{https://www.archivematica.org/}} and OpenRefine\footnote{\url{https://openrefine.org/}}.

In some cases, the metadata acquisition workflow with the aforementioned DLS' is complemented by using Microsoft tools, including MS Excel. \citet{jan2018application} demonstrates the use of Excel for structuring and validating heritage resource data, such as site IDs, names, geometries, and descriptions, before integrating the data into advanced platforms like Arches-HIP for geospatial analysis and management. Moreover, \citet{Kuzma2020} mention the use of Excel to process metadata of 35,092 topographic maps from the National Library of Poland. The precise role of Excel in this workflow, however, is not clear, as the authors only mentioned that ``scripts and formulas were defined to classify and analyze obtained data''~\citep[p. 5]{Kuzma2020}. Besides these minor uses of MS Excel, no work has been found containing a thorough description of a metadata acquisition workflow with Microsoft 365 cloud package tools on the forefront.

Utilizing Excel spreadsheets for metadata collection offers the benefit of a user-friendly interface, reducing the learning curve for participants. This advantage is particularly valuable in projects lacking designated domain experts, where individuals are enlisted on an \emph{ad hoc} basis for specific tasks as they arise during project execution. In such scenarios, the familiarity of Excel's interface enhances efficiency by enabling quick adaptation and seamless participation, even among those with varied levels of technical expertise. One way shown by~\citet{jan2018application} to ensure and enhance the quality of data entered in spreadsheets is by using Python scripts to extract, validate and integrate it with other resources such as ontologies, or another type of shared vocabularies.

Finally, \citet{kovalenko2013towards} discussion of tools for ontology population using spreadsheet data already shows some evaluated alternatives to the transformation of the collected metadata. In a more recent approach, \citet{denisova2022ontology} present the OntoGen tool for creating ontologies by transforming spreadsheet data. Another tool worth mentioning is Owlready, a Python package that provides convenient manipulation and population of OWL ontologies \citep[cf.][]{lamy2017owlready}.

Concluding, the review of existing metadata acquisition softwares and workflows shows that the main functionalities present in such systems are the support customizable metadata standards, such as Dublin Core and RDF, including validation for different datatypes. They also provide user-friendly interfaces for manual and automated metadata entry. Furthermore, some systems incorporate Linked Data principles, enabling metadata enrichment. In the following sections we describe our solution for offering similar features on a Microsoft cloud-based environment.

\section{Objectives and requirements}
\label{sec:goal}

Our ultimate goal is to create a knowledge base that adheres to FAIR principles~\citep{wilkinson2016fair}, that is linked to general knowledge bases---such as Wikidata~\citep{erxleben2014wikidata}, Geonames\footnote{\url{https://www.geonames.org/}.}, and YAGO~\citep{rebele2016yago}---as well as to other cultural heritage bases---including the pan-European Europeana project~\citep{haslhofer2011europeana} and bases describing the specific collections, e.g., the Kalliope catalog\footnote{\url{https://kalliope-verbund.info/en/}.} of approximately 600,000 records from 950 institutions located in German-speaking countries---and that will enable and facilitate a wide range of digital humanities research, including tasks done or supported by artificial intelligence tools.

As highlighted in the introduction, the objective of the work reported in this paper was to develop tools for metadata acquisition by domain experts and to combine them with existing systems and tools (described in Sect.~\ref{sec:ju}) into a unified digitization workflow for cultural heritage collections.
From the very beginning, the subset of collections stored in the Jagiellonian Library was chosen as the playground for the work.
Therefore, our research concerns practical challenges of massive digitization (hundreds of thousands of documents) of manuscripts of immense historical value for the purpose of a digital library.

The workflow needed to reach a balance between the two contradictory requirements: on one hand, it had to guarantee the high quality of metadata gathered by the domain experts; on the other, it should allow for relatively quick data collection and processing to ensure that the digitization process progresses smoothly and can be completed within a reasonable time frame.

The work on the prototypes was done in an iterative manner, allowing the needs, requirements, and limitations to be determined on an ongoing basis.
Among the most important of these were:
\begin{enumerate}
    \item Tight time frame for tool development -- four months were allocated for this task to be aligned with current digitization projects at Jagiellonian Library,
    \item Need for using tools that are as familiar as possible and/or easily accessible to domain experts -- i.e., typical office tools; due to the digital exclusion of some members of the research team, it was not possible to propose more complex solutions,
    \item Lack of human and financial resources to develop dedicated software,
    \item Lack of dedicated infrastructure to operate and maintain the tools,
    \item Preference for a web-based solution because domain experts use their private laptops, so we have no control over hardware and software setup, and can only force them to use a specific web browser.
\end{enumerate}

\section{Methods and tools}
\label{sec:methods}

Taking into account all the goals, requirements, and limitations outlined in Sect.~\ref{sec:goal}, we decided to use a generic tool such as a spreadsheet and develop a solution in the Microsoft 365 cloud, which is familiar to JU employees from their daily work.
To reduce the risk of errors and make such a tool easier to use, it was decided to develop a number of Office Script and Python scripts.
The whole is combined into a workflow (see Sect.~\ref{sec:workflow}) that uses the dedicated folder structure in Microsoft Sharepoint.
More specifically, the developed solution consists of the following:
\begin{enumerate}
    \item The Knowledge Matrix (TKM) -- an MS Excel file (see Sect.~\ref{sec:excels}),
    \item Standard Entries Catalog (SEC) -- a set of 4 MS Excel files (see Sect.~\ref{sec:excels}),
    \item The Mapping system (MAP) -- a sheet in TKM file for mapping between specific pages and actual scans (see Sect.~\ref{sec:excels}),
    \item Office Script that handles the buttons in TKM and performs initial validation (see Sect.~\ref{sec:validation}),
    \item Office Script to assist in adding new SEC records,
    \item A set of Python scripts performing more detailed validation of metadata entered into the TKM and SEC (see Sect.~\ref{sec:validation}),
    \item The Arcarium Ontology (TAO) -- interoperable data model for storing and retrieving rich metadata of manuscript collections housed at JU (see Sect.~\ref{sec:onto}),
    \item A set of Pythons scripts for populating knowledge stored in TKM and SEC into TAO (see Sect.~\ref{sec:onto}),
    \item A set of documents including: a description of the workflow, instructions for using TKM and SEC, a file for reporting errors and comments by domain experts.
\end{enumerate}

\subsection{Designing TKM, SEC, and MAP}
\label{sec:excels}

\begin{figure}[p]
    \centering
    \includegraphics[width=.95\textwidth]{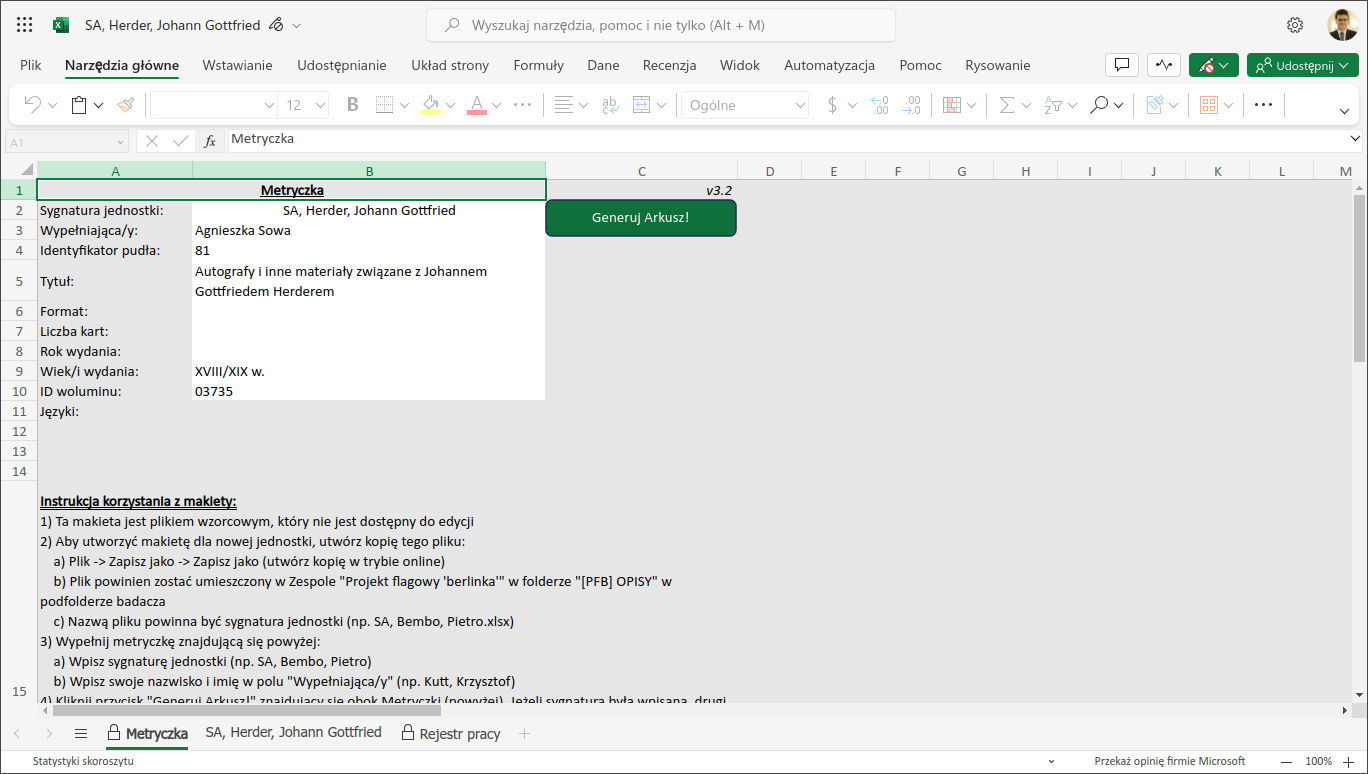}
    \caption{Metric sheet of the TKM containing general information of the whole unit (in Polish), i.e., the metadata that will be converted to MARC 21 format to create records in library catalog (title, card number, format, etc).}
    \label{fig:tkm-metric}
\end{figure}

\begin{figure}[p]
    \centering
    \includegraphics[width=.95\textwidth]{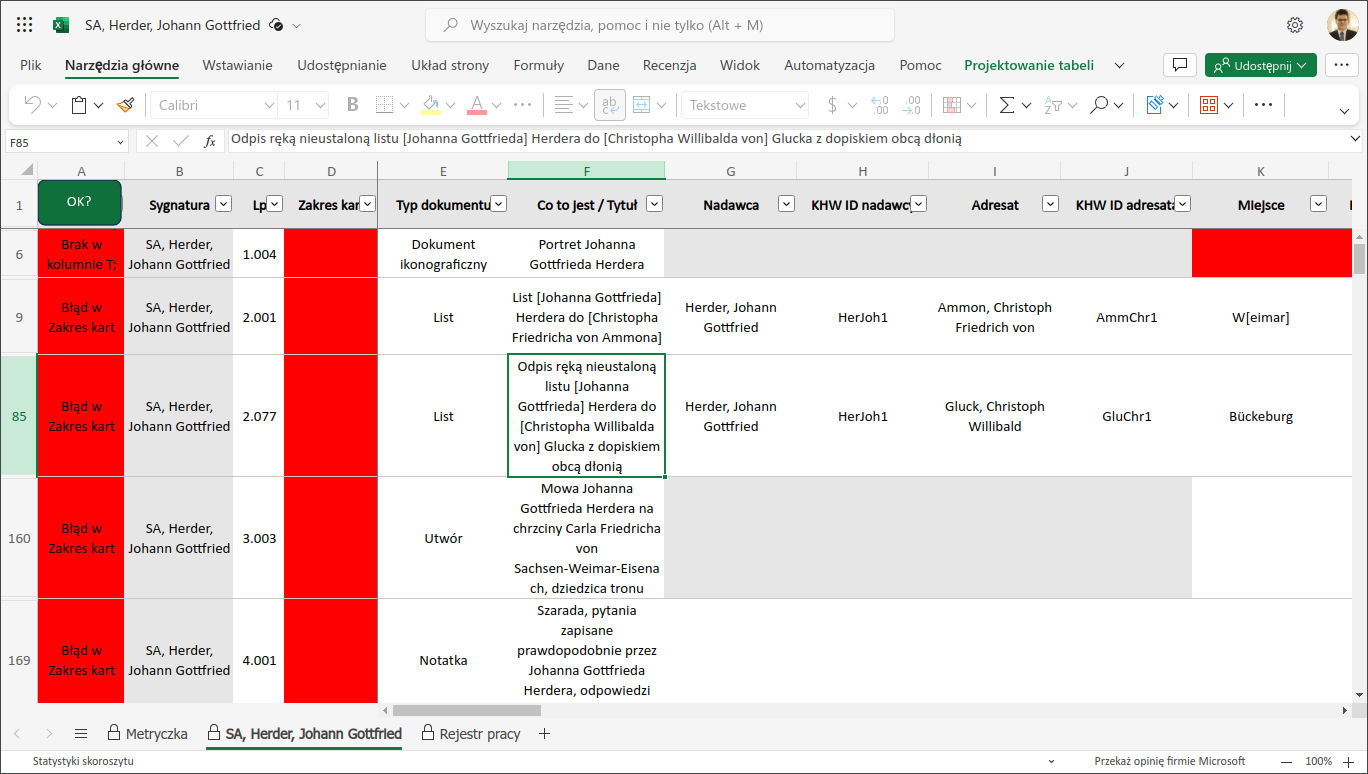}
    \caption{Main sheet of the TKM with one row per document in a given unit (in Polish). One can see the result of the Office Script (errors marked with red color, fields not used in a given category marked with gray background) and the ``OK?'' button that triggers the script.}
    \label{fig:tkm-documents}
\end{figure}

The Excel-based form for entering document metadata is the key data structure.
First, it was assumed that each unit\footnote{A set of documents marked with one shelfmark; in collection used for pilot study, one shelfmark was associated with all documents related to individual person.} would correspond to one file, in which one of the sheets would be a metric containing the description of a unit as a whole (see Fig.~\ref{fig:tkm-metric}), and the other would be used to collect metadata of individual documents represented by subsequent rows (see Fig.~\ref{fig:tkm-documents}). In turn, columns of the latter sheet would represent successive fields in the document description. Thus, The Knowledge Matrix emerged.
It was decided that each document would be classified into one of nine categories, a decision that determines the assignment of a particular set of fields to it.
The list is a slightly modified version of the scheme used for the Varnhagen Collection~\citep{kitahuber2022varnhagen} and comprises the following categories:
\begin{enumerate}
    \item Portraits, images, and drawings,
    \item Outgoing correspondence,
    \item Creative works (literary and other),
    \item Personal materials,
    \item Historical materials, diplomas,
    \item Printed materials and press clippings,
    \item Incoming correspondence,
    \item Foreign materials (notes about individuals, various letters),
    \item Library materials (covers, bookmarks, historical catalog cards).
\end{enumerate}

While the structure of the metric is very strict, as it contains fields necessary for creating records in the Alma library catalog (after conversion to MARC 21),
the metadata collected for each document will be stored exclusively in the ontology and ultimately in a dedicated system that will enable their viewing.
Therefore, it is more flexible and can be adapted to the specific nature of the resources being described.
In the pilot study described, the metadata structure was created from scratch by a team of domain experts responsible for describing the documents (the whole data model can be explored in the shared ontology, see the Data availability section).
Some basic fields are ascribed to all document types, e.g., ``what is it / document title'', while other are associated with particular categories, e.g., ``sender'' and ``recipient'' are assigned to correspondence (categories 2 and 7), ``issuer'' is assigned to official documents (categories 4 and 5), and ``author'' to creative works (category 3).
The description fields were further divided into mandatory (as ``document number'') and optional (as ``date remarks'').
Detailed content restrictions were imposed on some of the fields, e.g., ``date'' field could only accept one of the specified date formats.
The most important from the validation point of view (see Sect.~\ref{sec:validation}) is ``document number'' filled with a string in format $x.y$, where $x$ is a digit of encoding document category of 1 to 9, and $y$ is a sequential number of a document within a unit.

\begin{figure}[t]
    \centering
    \includegraphics[width=\textwidth]{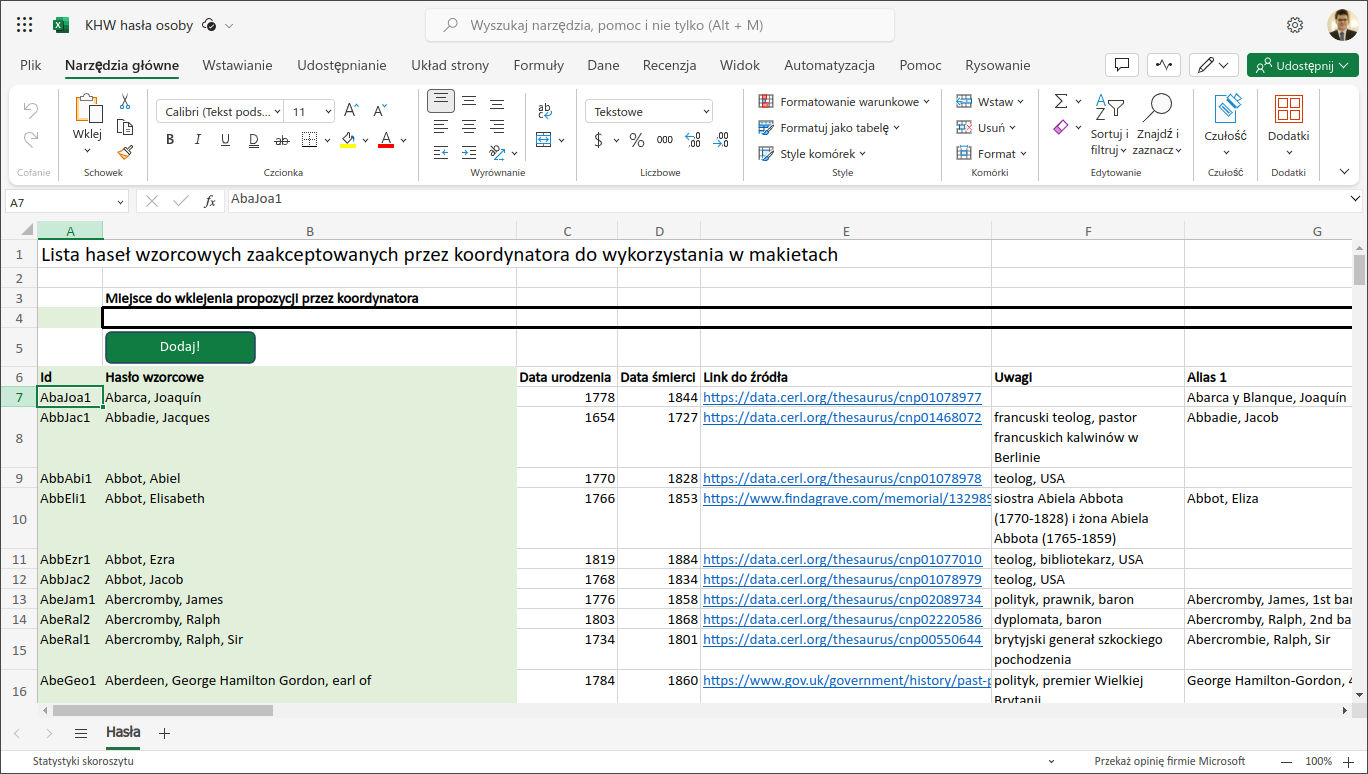}
    \caption{SEC sheet with person-related metadata (in Polish).}
    \label{fig:sec}
\end{figure}

During subsequent iterations, the need for separate acquisition of person- and place-related metadata appeared. The problem arose from the fact that numerous persons and geographical objects were referred to in the documents under different names, resulting from variant spellings across languages or from the fact that a person or institution could bear different names at different stages of their existence. This need led to the creation of the Standard Entries Catalog.
The SEC consists of four Excel files.
Two of them are the parts of the actual catalog, presenting information about persons (see Fig.~\ref{fig:sec}) and locations, respectively.
Only the SEC coordinator is able to enter new data or modify existing entries, while other experts have read-only access.
The other two files serve as collectors of proposals from domain experts: they can write suggestions of new entries or extending or modifying the existing entries in both main SEC files.
The TKM records link to specific SEC entries by providing the SEC ID in a dedicated column.
The SEC records are linked to external databases by providing URLs (e.g., Geonames for geographic names).

Finally, when the TKM file is finished, the validation script (see Sect.~\ref{sec:validation}) extends it automatically with a MAP sheet designed to store bindings between unit's card numbers and their scans.
A single page may sometimes require more than just two scans (one for its recto, another for its verso side).
There may be one or more paste-in slips attached to the page that require separate scanning.
A telling example of this problem is one of the pages from Alexander von Humboldt's Kosmos manuscript whose scanning produced as many as 14 different files~\citep[see pp. 62-75 in][]{humboldt1845kosmos}.
Therefore, to enable linking particular documents with their scans, a system of mapping card numbers to scan files---done manually by librarians in a separate MAP sheet---and mapping units with their documents to specific folder in a folder tree where the files are stored was introduced.

\subsection{Data validation scripts}
\label{sec:validation}

Data entered through interfaces (Excel files) must be evaluated to ensure the quality of the metadata is as high as possible.
Two programs that perform data validation have been developed for this purpose: 1) \emph{Online validator}, which validates data \emph{on the fly}, as they are entered by the researcher into the TKM sheet (via an Office Script), and 2) \emph{Offline validator}, which do its work after the sheet has been filled out completely -- in this case, a full analysis is possible, which is technically impossible to perform at the Excel online script level (via a set of Python scripts).
In both cases, the syntax of the input data and its completeness are analyzed.
\emph{Offline validator} additionally performs limited semantic validation including, but not limited to: verification whether the Author's life span is not greater than 110 years, whether the entered dates actually existed in the calendar, whether the entered SEC IDs indeed exist in the SEC and correspond to the specified person/place.
Feedback is provided to users through appropriate messages in a dedicated column A and cell background colors (in \emph{Online validator}) and in the form of a report in a PDF file (in \emph{Offline validator}).

Both programs were dedicated to data validation, however, during the course of work, their scope of operation was extended with additional functionalities resulting from the current needs:
\begin{enumerate}
    \item Forcing the process of filling out the TKM (the metric is filled out first, and then the main sheet);
    \item Blocking fields unavailable for a given category of document (e.g., blocking of the ``sender'' field for documents in the ``Portraits, images, and drawings'' category) in order to facilitate the work of users;
    \item Updating the TKM with new columns and fields, performed by \emph{Online validator} (e.g., adding new fields that were not provided for earlier; the script performs this change without the need for additional user intervention);
    \item Validation of SEC entries by \emph{Offline validator} (an independent report is generated, visible to the SEC coordinator).
\end{enumerate}

For \emph{Offline validator}, a dedicated virtual server on the JU infrastructure was set up and configured to perform a synchronization with Sharepoint once a day, run the validation scripts, and perform the synchronization again to propagate the validation reports.

\subsection{TAO ontology population}
\label{sec:onto}

The final step of the entire workflow is the ontology population, i.e., information about documents and individuals entered through the TKM and SEC interfaces is translated using custom scripts based on the owlready2 library into a knowledge graph in RDF notation and placed as instances in the TAO ontology. The result is a knowledge base built with rich metadata that allows querying in SPARQL to obtain answers to, e.g., (a) ``To what shelfmark does a specific document belong?'', (b) ``In what languages are the documents in a given shelfmark?'', (c) ``Does a given person appear in the collection under other names, and if so, which ones?''.
A detailed description of ontology design is presented in~\citep{lvm2024tao}.

\section{Evaluation}
\label{sec:workflow}

Data entry and validation are only the initial steps in an overall metadata acquisition process for digitization integrated with JL's procedures.
Working closely with JL, through a series of iterations, existing processes at JL were first identified, and then a new complete process was developed using Excel input interfaces, the Microsoft 365 cloud, and JL's existing infrastructure.
Importantly, the process was subjected to practical evaluation through two pilots and two workshops with international guests, which allowed for its improvement, and confirmation of its correctness and usability for JL.

Designing, organizing, and implementing the workflow process, and consequently its implementation in practice, was a major organizational challenge due to three key factors. Firstly, it included the activities of three different groups of people, partly working in different places, with different levels of IT competences, using different IT tools on a daily basis and coming from different work and organizational cultures: (a) domain experts -- humanities researchers, (b) computer scientists -- researchers, academic teachers, doctoral students and practitioners, and (c) library workers.
Secondly, the workflow was to be fed with different resources at different times: on the one hand, original manuscripts that require special protection and security procedures, and on the other hand, two types of digital documents: Excel files filled with metadata and scans of the manuscripts.
Thirdly, the Jagiellonian Library did not have a ready-made tool capable of comprehensively handling such a workflow. As indicated earlier, it was decided to use the Sharepoint platform offered in the cloud version of MS Office available to all University employees, but also to a certain extent the Alma library system and folder structures opened for this purpose on a dedicated server NAS (Network Attached Storage), as well as a paper register to control the transfer of original documents.

\begin{figure}[th]
    \centering
    \includegraphics[width=.8\textwidth]{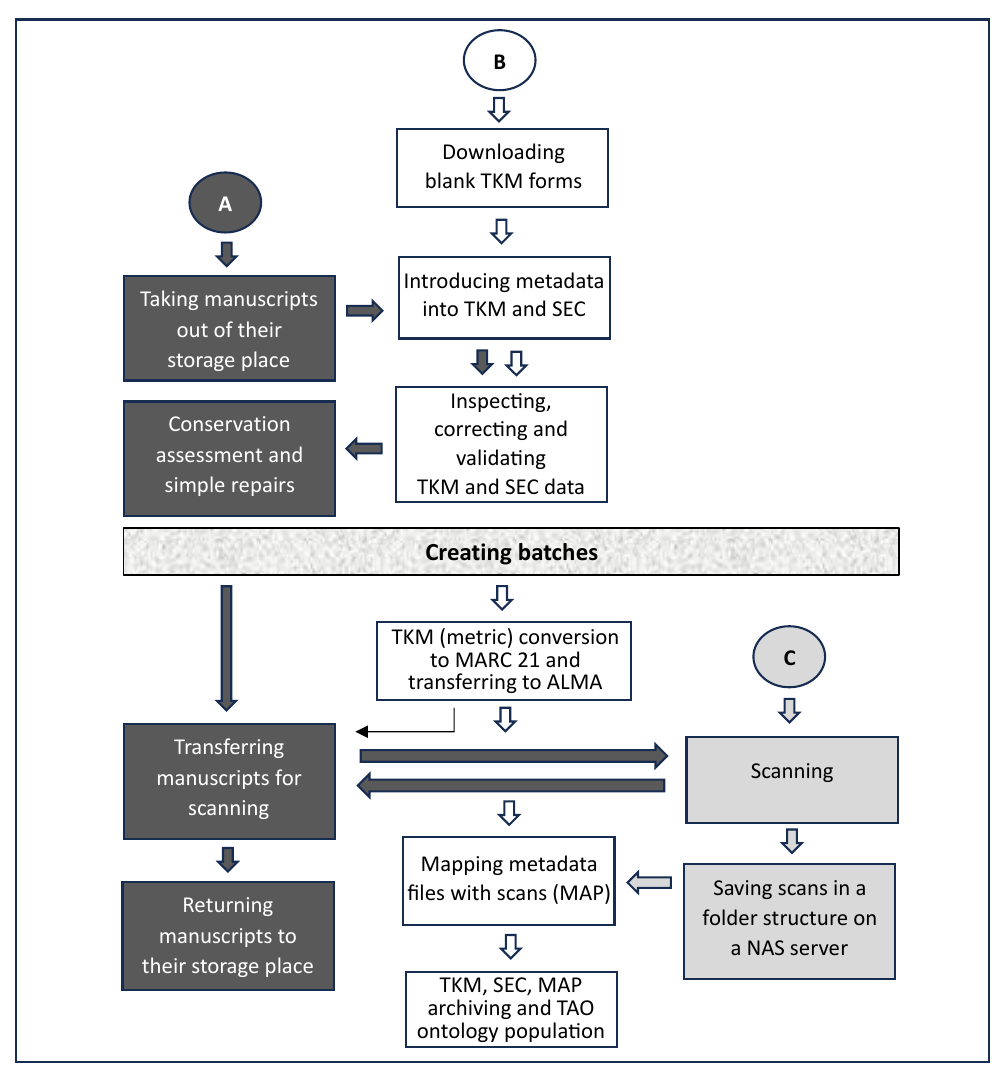}
    \caption{The digitzation workflow in practice.}
    \label{fig:workflow}
\end{figure}

In practice, a bundle of three parallel paths was de facto formed, connecting with each other at specific points (which is presented in a simplified and synthetic way in Fig.~\ref{fig:workflow}).
Path ``A'' refers to the handling of original manuscripts, the transfer of which is controlled analogously using paper documentation containing signatures confirming the transfer and receipt of objects by the people involved.
Path ``B'' is the main axis of the entire project -- it refers to the ``life'' of Excel files and intended for collecting and storing (at this stage of the project) metadata describing individual manuscripts. This path is implemented entirely in the digital environment within the structure of folders created on a dedicated Sharepoint website.
Path ``C'' includes activities related to the scanning process and is closely related to both paths mentioned above.

A practical assessment of the workflow has led to work being performed partly in a ``batch mode''. The flow is paused at the place marked in Fig.~\ref{fig:workflow} as ``Creating batches'', a larger number of finally accepted files with metadata are collected and only such a ``batch'' is processed further. It is then processed as one action, after which work in this part of the workflow is suspended and waiting for the next ``batch''. At the same time, in the first part of the workflow, which refers to the creation of metadata, work progresses continuously.

The bundle is completed by the execution of scripts that, based on archived metadata files, populate instances in the TAO ontology. During the pilot studies, this step was run manually, but ultimately it is planned to have routines that will call this step regularly to automatically populate the knowledge base with newly acquired metadata.

In the two pilot studies, the researchers created 235 TKMs and a number of records of individuals and places in the SEC.
As the pilot studies were evaluating the continuous work of collecting metadata, according to the above workflow, their termination at a particular point in time meant that many TKMs were left in a draft version, which may contain errors or omissions.
Therefore, during the ontology population step, it was assumed that any records with errors are skipped and errors are not analyzed, as they are mostly due to the specifics of the pilot studies (in the actual workflow, ontology population step will be done only on finished and validated TKMs).

As a result, a knowledge base consisting of 6084 axioms describing 852 individuals (people, documents, \ldots) was created to evaluate the capabilities of the developed data model. The ontology has been made available in a public git repository available at
\url{https://gitlab.geist.re/pro/tao},
along with a basic user manual to help familiarize oneself with the data we have collected.
A detailed description of the data model can be found in~\citep{lvm2024tao}.

Evaluation of the entire workflow during two pilots and in two workshops with international guests confirmed the usefulness of the proposed solution.
It allowed efficient collection of rich metadata by people with diverse backgrounds and integration with already existing digitization processes.
The resulting knowledge base makes it possible to obtain answers to queries according to a pre-developed set of competency questions~\citep[cf.][]{lvm2024tao}.

\section{Conclusions}%
\label{sec:conclusions}

When joining as a group of knowledge engineers and computer scientists in cultural heritage initiatives at the Jagiellonian University, we hit upon the challenge of preparing tools to collect metadata in a short period of time, with very limited human and financial resources and for a group of domain experts, some of whom are digitally excluded.
It was important to keep in mind the scalability of the solution to hundreds of thousands of manuscripts and to take care of the validation of the metadata to be able to create a functional knowledge graph-based library out of it in the future.
To achieve this, we designed a solution based on Microsoft 365 cloud, using MS Excel files as data entry interfaces, Office Script and Python scripts for data validation, a dedicated folder structure in MS Sharepoint for file storage, and a set of Python scripts for ontology population, and then together with the Jagiellonian Library, we formulated a workflow that combines existing systems and procedures with the proposed tools.
The workflow has already been evaluated in two pilot studies and in two workshops with international guests, which allowed for its refinement and confirmation of its correctness and usability for JL.

\section*{Acknowledgement}

We would like to express our gratitude to Remigiusz Sapa---Director of the Jagiellonian Library---for overseeing all the efforts carried out at the JL to support the design and validation of the presented workflow, and to all the JL staff involved in these activities.

\section*{CRediT authorship contribution statement}

\textbf{Krzysztof Kutt}: Writing -- original draft, Writing -- review \& editing, Software, Methodology, Data curation, Conceptualization.
\textbf{Luiz do Valle Miranda}: Writing -- original draft, Writing -- review \& editing, Software, Validation.
\textbf{Jakub Gomu\l{}ka}: Writing -- original draft, Methodology, Conceptualization.
\textbf{Grzegorz J. Nalepa}: Writing -- review \& editing, Supervision, Resources, Project administration, Conceptualization, Funding acquisition.

\section*{Funding sources}

This publication was funded by a flagship project ``CHExRISH: Cultural Heritage Exploration and Retrieval with Intelligent Systems at Jagiellonian University'' under the Strategic Programme Excellence Initiative at Jagiellonian University.
The research has been supported by a grant from the Priority Research Area (DigiWorld) under the Strategic Programme Excellence Initiative at Jagiellonian University.

\section*{Declaration of generative AI and AI-assisted technologies in the writing process}

During the preparation of this work the authors used Writefully in order to improve the readability and language of the manuscript. After using this tool, the authors reviewed and edited the content as needed and take full responsibility for the content of the published article.

\section*{Declaration of competing interest}

The authors declare that they have no conflict of interest.

\section*{Data availability}

The data model developed from the data collected in the pilot studies was made available in a publicly available git repository at:
\url{https://gitlab.geist.re/pro/tao}.

\section*{Code availability}

All software supporting the workflow (Office Script scripts, a collection of Python scripts) has been created under the GNU GPL v3.0 license and is available upon e-mail request sent to the authors of the paper.

\bibliographystyle{elsarticle-harv}
\bibliography{../geistbib/culheripub,../geistbib/culheriteam}

\end{document}